# Composition and hydrogen storage structure of $Ti_2CT_x$ MXene with ultrahigh hydrogen storage capacity

Sen Jin, Qianku Hu*, Aiguo Zhou*

*School of Materials Science and Engineering, Henan Polytechnic University, Jiaozuo, Henan 454003, People's Republic of China*

\* Corresponding authors: hqk@hpu.edu.cn, zhouag@hpu.edu.cn

Recently, Liu et al. reported that $Ti_2CT_x$ MXene have ultra-high hydrogen storage capacity (8.8 wt.%) at room temperature [1]. For the purpose to clearly understand the hydrogen storage (H-storage), the composition of studied samples should be clearly characterized and the H-storage structure need be explored. To achieve 8.8 wt.% capacity, 3 layers of $H_2$ molecules need be stored in the interlayer space of MXene with the structure of $Ti_2CF_2H_{14}$. The $H_2$ layers with graphene-like 2D structure are in solid/liquid state at room temperature, which is significant in the explore new materials with surprising properties.

Liu's work[1] is the first experimental report on MXenes' excellent H-storage performance, though it has been theoretically predicted in 2013 [2]. After the first theoretical prediction, there are more theoretical works [3,4,5], and some experimental works on MXenes' adsorption of $CH_4$ [6] and $CO_2$ [7]. However, there are no experimental reports on the hydrogen adsorption storage of MXenes, except the recent work by Liu et al. [1]. This is because that only MXenes with special microstructure can show high hydrogen storage capacity, as clarified in that research [1].

MXene is a novel two-dimensional transition metal carbide/nitride prepared by etching A element from MAX phase (a family of compound with layered structure) [8,9,10]. The surface of MXenes prepared in HF etching solution is always terminated with -F, -OH, and/or -O [11,12]. Hydrogen is adsorbed on the terminations (F/OH/O), and stored in interlayer space of MXene stacks. In the work of Liu et al. [1], the preferred termination is F and the preferred interlayer-distance is ~7 Å.

The composition of Ti$_2$C MXene were characterized by X-ray diffraction (XRD) in Ref. 1. The XRD pattern is shown in Fig. 1a, which is copied from Fig. 1c of Ref. 1. Some peaks are labeled by the numbers (1-6) in purple color.

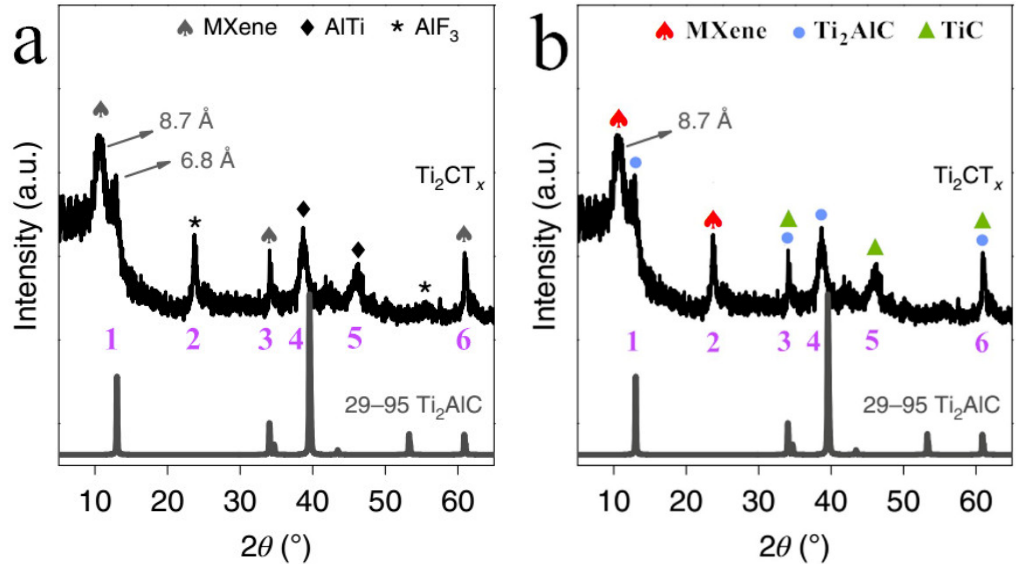

**Figure 1.** (a) XRD pattern of as-prepared Ti$_2$C MXene with the reference of Ti$_2$AlC, copied from Ref. 1. Some peaks are labeled by the numbers (1-6) in purple color, (b) XRD pattern indexed according to the discussion in this paper

In Fig. 1a, peak 4 and peak 5 were indexed as AlTi's peak. However, it is not reasonable to generate AlTi, an intermetallic compound by the reaction between HF and Ti$_2$AlC. Thus, these two peaks may belong to other matters, such as Ti$_2$AlC and TiC. Ti$_2$AlC is the initial material to make Ti$_2$C MXene. In Ref. 1, Ti$_2$AlC was "incompletely etched" to make Ti$_2$C MXene. Thus it is very possible that some residual Ti$_2$AlC exist in the final samples. The pattern of Ti$_2$AlC is shown in the bottom of the figure. Almost all peaks of Ti$_2$AlC can be found in the Ti$_2$CT$_x$'s pattern (peak 1, peak 3, peak 4 and peak 6), except that peak 4 shifts to low angle a little. This shift may be resulted by the crystal deformation due to etching. Thus, peak 4 may be the peak of "incompletely etched" Ti$_2$AlC. Moreover, TiC is a common impurity in the synthesis of titanium contained MAX phases and MXenes [13, 14]. The characteristic peaks of TiC are at 2θ=35.9°, 41.7° and 60.4°. Peak 3, peak 5 and peak 6 have the 2θs close to these values. Considered the crystal deformation due to etching, peak 5 may belong to impurity TiC.

In Fig. 1a, Peak 2 (2θ=~23°) was indexed as $AlF_3$'s peak. The other peak indexed as $AlF_3$'s peak (2θ=~56°) is very weak and can be neglected as background noise. From PDF cards of $AlF_3$ (43-0435, 44-0231 and 47-1659), more peaks should appear if $AlF_3$ existed. Thus the only $AlF_3$ peak may be the peak of other matters. It is noted that 2θ of peak 2 (~23°) is two times of the 2θ of MXene's (002) peak (~11°). Thus, this peak should be the (004) peak of MXene as indexed in other research of MXene[15].

According to above discussion, the figure was revised and shown in Fig. 1b. The compositions of the samples should be MXene with residual $Ti_2AlC$ and TiC impurity, rather than MXene with AlTi and $AlF_3$.

In Ref. 1, density functional theory was used to calculate the H-storage structure. The optimal structure is shown in Fig. 2a, which is copied from Supplementary Fig. 22 of Ref. 1. As shown in Fig. 2a, 2 layers chemically bonded H atoms (light blue balls) and 1 layer weak chemically absorbed $H_2$ molecules (brown balls) are adsorbed in the interlayer space between two $Ti_2C$ layers. The H-storage structure is $Ti_2CH_4$. If both chemically bonded H and weak chemically absorbed H are considered as stored H, the theoretical capacity is 3.7 wt.%. Moreover, the chemically bonded H is very difficult to be released. Thus, the theoretical capacity should be much lower than 3.7 wt.%. However, the measured capacity was 8.8 wt.%. Therefore, that H-storage structure in Fig. 2a cannot explain so high H-storage capacity. New H-storage structure should be explored.

The surface of $Ti_2C$ MXene is always terminated with F/OH/O and only $Ti_2C$ MXene with F termination show ultrahigh H storage capacity in Ref. 1. Thus, $Ti_2CF_2$ rather than $Ti_2C$ should be used as substrate to adsorb and store H. If the theoretical capacity is assumed to be the highest measured capacity (8.8 wt.%), the number of H atoms adsorbed by $Ti_2CF_2$ unit cell was calculated to be ~14. Then the theoretical H-storage structure should be $Ti_2CF_2H_{14}$. To achieve 8.8 wt.% capacity, 14 H atoms or 7 $H_2$ molecules need to be adsorbed by one $Ti_2C_2F_2$ unit cell and stored in the interlayer space. It is necessary to know the arrangement of the 14 H atoms.

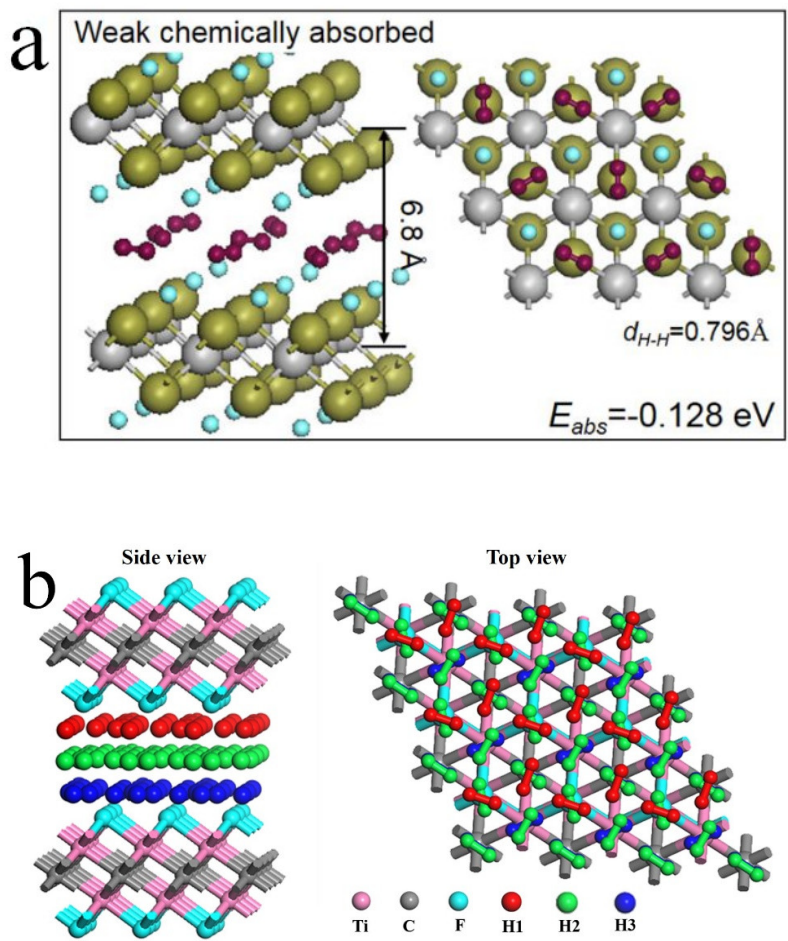

**Figure 2.** (a) Crystal structure (side view and top view) of $Ti_2C$ stack with chemically bonded H atoms and weak chemically bonded $H_2$ molecules ($Ti_2CH_4$) for DFT calculations in Ref. 1, (b) Crystal structure of $Ti_2CF_2$ stack with three layers of weak chemically bonded $H_2$ molecules ($Ti_2CF_2H_{14}$) based on the discussion in this paper

As shown in Fig. 2a, only the H directly bonded with Ti atoms exist as atoms (first layer and third layer, light blue balls). Thus all H adsorbed on $Ti_2C_2F_2$ should be weak chemically bonded $H_2$ molecules. As shown in Fig. 2b, we arrange 7 $H_2$ molecules in the interlayer space of $Ti_2CF_2$. The 7 $H_2$ can be divided into 3 layers. Red layer has 2 $H_2$, green layer has 3 $H_2$ and blue layer has 2 $H_2$. This is a reasonable H-storage structure that can perfectly explain the ultrahigh capacity measured in the experiments. If this structure is true, the distance between two nearby $H_2$ molecules is very short. The three layers of $H_2$ molecules are in solid/liquid state rather than gas state. And the $H_2$ structure is hexagonal 2D structure, similar with the graphene structure. This is a possible way to make "hydroene", a new 2D material with the composition of $H_2$ and graphene-like 2D structure. Of course, more theoretical and experimental works are required to validate the H-storage structure.

In summary, we give a new XRD analysis and propose a new H-storage structure that can rationally explain the ultrahigh hydrogen storage performance of $Ti_2C$ MXene. The new H-storage structure open a door to explore new hydrogen material in sold/liquid state at room temperature. The new "hydroene" is hydrogen with graphene-like 2D structure.

## Competing interests statement

The authors declare no competing interests.

## Acknowledgements

This work was supported by National Natural Science Foundation of China (51772077), Program for Innovative Research Team (in Science and Technology) in the University of Henan Province (19IRTSTHN027), Fundamental Research Funds for the Universities of Henan Province (NSFRF200101).